\documentclass[prd,aps,amsfonts,showpacs,longbibliography,notitlepage,twocolumn,groupedaddress,nofootinbib,superscriptaddress]{revtex4-1}
\usepackage{preamble}

\begin{document}
\title{Slow mixing of Gibbs sampling for the quantum Curie-Weiss model}

\author{Xiaoyang Huang}
\email{xhuang1@perimeterinstitute.ca}
\affiliation{Perimeter Institute for Theoretical Physics, Waterloo, ON N2L 2Y5, Canada}

\begin{abstract}
    Hamiltonian evolution can spread local operators across a dense quantum magnet. We show
that this global support does not remove the bottleneck to thermalization in the quantum Curie–Weiss model: at fixed low temperature and weak transverse field, the Chen–Kastoryano–Gily\'en Gibbs sampler [arXiv:~2311.09207] has an exponentially large mixing-time lower bound. A dynamical quantum bottleneck theory that generalizes the classical notion of conductance has been proposed alongside existing frameworks that rely on spatial locality. We also give a complementary interpretation of slow mixing using an effective Fokker-Planck equation.
\end{abstract}

\date{\today}
\maketitle

\section{Introduction}

Sampling from Gibbs states is an important tool for studying statistical many-body systems at finite temperature. It is widely used in condensed matter, quantum information, and machine learning, with applications including extracting correlation functions to probe thermal phases of matter \cite{Temme2011Metropolis}, preparing asymptotically low-temperature states by quenching the temperature \cite{TerhalDiVincenzo2000Equilibration}, and training (quantum) Boltzmann machines for machine learning \cite{Amin2018QBM}. Unlike its classical counterpart, an efficiently implementable quantum Gibbs sampling algorithm requires a careful balance between spectral resolution and simulation time because of the uncertainty relation; substantial progress has been made only recently \cite{Chen2025Thermal}. In particular, it has been shown that an efficient sampling algorithm generated by a Lindbladian satisfying quantum detailed balance can be explicitly constructed for generic local quantum Gibbs states \cite{ckg,Ding-lin-lin,Gilyen2024GlauberMetropolis}.
% CITATION-SCOPE: Metcalf2020Cooling is related engineered-cooling background; the exact temperature-quench wording is not verified. CKG implementation efficiency remains conditional on mixing time.

Mixing time is crucial for estimating the cost of Gibbs sampling. It describes how quickly a sampling algorithm can bring an arbitrary initial state to the target Gibbs state with a fixed accuracy. An important property of the mixing time is its system-size dependence. For example, when the quantum Gibbs state realizes certain non-trivial ``orders'', the mixing time will be exponentially large in the system size, indicating slow mixing, because otherwise the correlations are likely to cluster \cite{KastoryanoEisert2013}. 
% CITATION-SCOPE: KastoryanoEisert2013 supports rapid-mixing-to-clustering under locality/dynamical hypotheses; it does not establish the generic exponential-mixing claim from order.
While considerable effort has been devoted to proving upper bounds on how the mixing time scales with the system size to demonstrate the low cost of sampling algorithms \cite{Temme2010Chi2,KastoryanoTemme2013LogSobolev,KastoryanoBrandao2016,Bardet2023Rapid,Rouze2026Thermalization,Rouze2026Optimal,Bergamaschi2026AllToAll,BergamaschiGheissariLiu2026LogicalSector}, only a few works are devoted to providing lower bounds \cite{GKZ2024,RPBK2026}.

% Add these lines to the preamble of your manuscript:
% \usepackage{amsmath,graphicx,tikz}
% \usetikzlibrary{arrows.meta}
%
% Then paste or \input this file in the manuscript body.
% The drawing is designed at 8.55 cm wide and scales to \columnwidth.

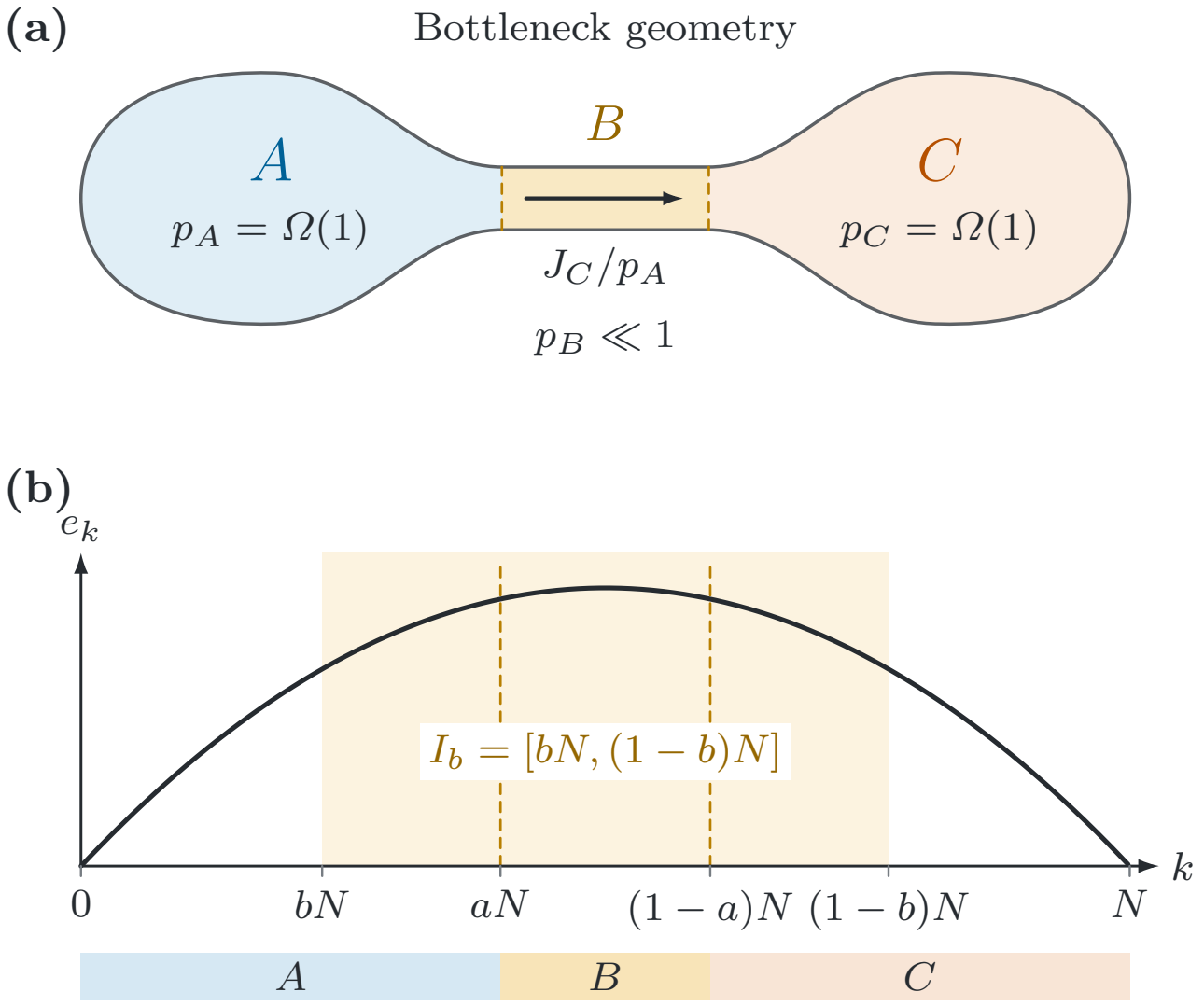
\begin{figure}[t]
  \centering
  \begingroup
  % BEGIN FIGURE DEFINITIONS
  \definecolor{bnA}{HTML}{0072B2}
  \definecolor{bnB}{HTML}{E69F00}
  \definecolor{bnC}{HTML}{D55E00}
  \definecolor{bnInk}{HTML}{262B30}
  \definecolor{bnMuted}{HTML}{777D83}
  % END FIGURE DEFINITIONS
  \resizebox{\columnwidth}{!}{%
  % BEGIN TIKZ
  \begin{tikzpicture}[
    x=1cm,y=1cm,
    font=\fontsize{8}{9.5}\selectfont,
    text=bnInk,
    line cap=round,line join=round,
    axis/.style={draw=bnInk,line width=0.55pt,
                 -{Latex[length=1.5mm,width=1.0mm]}},
    cut/.style={draw=bnB!80!black,line width=0.5pt,
                dash pattern=on 2pt off 1.6pt},
    panel/.style={anchor=west,font=\bfseries\fontsize{9}{10}\selectfont}
  ]
  \path[use as bounding box] (0,0) rectangle (8.55,7.20);

  % ---------- (a) Bottleneck cartoon ----------
  \begin{scope}[yshift=0.18cm]
  \node[panel] at (0.04,6.78) {(a)};
  \node at (4.375,6.78) {Bottleneck geometry};

  % A broad pair of basins joined by a narrow passage.
  \def\bnOutline{%
    (0.70,5.60)
    .. controls (0.70,6.22) and (1.30,6.50) .. (2.03,6.48)
    .. controls (2.73,6.48) and (3.04,5.82) .. (3.65,5.82)
    -- (5.10,5.82)
    .. controls (5.71,5.82) and (6.02,6.48) .. (6.72,6.48)
    .. controls (7.45,6.50) and (8.05,6.22) .. (8.05,5.60)
    .. controls (8.05,4.98) and (7.45,4.70) .. (6.72,4.72)
    .. controls (6.02,4.72) and (5.71,5.38) .. (5.10,5.38)
    -- (3.65,5.38)
    .. controls (3.04,5.38) and (2.73,4.72) .. (2.03,4.72)
    .. controls (1.30,4.70) and (0.70,4.98) .. (0.70,5.60)
    -- cycle
  }
  \begin{scope}
    \clip \bnOutline;
    \fill[bnA!12] (0,4.6) rectangle (3.65,6.6);
    \fill[bnB!23] (3.65,4.6) rectangle (5.10,6.6);
    \fill[bnC!12] (5.10,4.6) rectangle (8.4,6.6);
  \end{scope}
  \draw[bnInk!75,line width=0.65pt] \bnOutline;
  \draw[cut] (3.65,5.38)--(3.65,5.82);
  \draw[cut] (5.10,5.38)--(5.10,5.82);

  % Representative configurations in the two basins.
  % \foreach \x/\y in {1.20/5.85,1.52/6.18,2.48/6.12,
  %                      2.91/5.75,1.16/5.29,1.55/5.03,2.50/5.06}{
  %   \fill[bnA!65] (\x,\y) circle[radius=1.15pt];
  %   \fill[bnC!65] ({8.75-\x},\y) circle[radius=1.15pt];
  % }
  \node[text=bnA!85!black,font=\fontsize{12}{13}\selectfont]
    at (2.03,5.87) {$A$};
  \node[text=bnC!85!black,font=\fontsize{12}{13}\selectfont]
    at (6.72,5.87) {$C$};
  \node at (2.03,5.38) {$p_A=\Omega(1)$};
  \node at (6.72,5.38) {$p_C=\Omega(1)$};
  \node[text=bnB!65!black,font=\fontsize{10}{11}\selectfont]
    at (4.375,6.13) {$B$};
  \node at (4.375,4.62) {$p_B\ll 1$};
  \draw[axis,line width=0.8pt] (3.82,5.60)--(4.93,5.60);
  \node at (4.375,5.12) {$J_C/p_A$};

  % \node at (4.375,4.30)
  %   {$\displaystyle
  %     t_{\mathrm{mix},\epsilon}\ge
  %     \frac{p_A}{J_C}\left(p_C-q_C-\frac{\epsilon}{2}\right)$};
  \end{scope}

  % ---------- (b) Curie--Weiss shell energies ----------
  \node[panel] at (0.04,3.73) {(b)};
  % \node at (4.375,3.73)
  %   {$\displaystyle e_k=\frac{2k(N-k)}{N},\qquad 0<b<a<\frac12$};

  % Illustrative coordinates; b<a<1/2.
  \def\bnXzero{0.70}
  \def\bnWidth{7.35}
  \def\bnYzero{1.10}
  \def\bnEscale{3.90} % y = y0 + Escale * (e_k/N)
  \def\bnBvalue{0.23}
  \def\bnAvalue{0.40}
  \def\bnEta{0.25}
  \pgfmathsetmacro{\bnXb}{\bnXzero+\bnWidth*\bnBvalue}
  \pgfmathsetmacro{\bnXa}{\bnXzero+\bnWidth*\bnAvalue}
  \pgfmathsetmacro{\bnXac}{\bnXzero+\bnWidth*(1-\bnAvalue)}
  \pgfmathsetmacro{\bnXbc}{\bnXzero+\bnWidth*(1-\bnBvalue)}
  \pgfmathsetmacro{\bnXend}{\bnXzero+\bnWidth}
  \pgfmathsetmacro{\bnYeta}{\bnYzero+\bnEscale*\bnEta}

  % The larger interval I_b contains the bottleneck sector B.
  \fill[bnB!12] (\bnXb,\bnYzero) rectangle (\bnXbc,3.30);
  \foreach \xx in {\bnXa,\bnXac}{
    \draw[cut] (\xx,\bnYzero)--(\xx,3.22);
  }
  \draw[axis] (\bnXzero,\bnYzero)--(8.26,\bnYzero)
    node[right,inner sep=2pt] {$k$};
  \draw[axis] (\bnXzero,\bnYzero)--(\bnXzero,3.30)
    node[above,inner sep=2pt] {$e_k$};

  \draw[bnInk,line width=0.95pt]
    plot[domain=0:1,samples=100]
      ({\bnXzero+\bnWidth*\x},
       {\bnYzero+\bnEscale*2*\x*(1-\x)});

  % Energy cutoff and the annotation supplied in the question.
  % \draw[bnA,dashed,line width=0.6pt]
  %   (\bnXzero,\bnYeta)--(\bnXend,\bnYeta);
  % \node[anchor=east,text=bnA,inner sep=1pt]
  %   at (0.58,\bnYeta) {$\eta N$};
  % \node[text=bnA,fill=bnB!12,inner sep=1.5pt]
  %   at (4.375,1.56) {$\eta<2b(1-b)+h$};
  % If matching Proposition 1 in the attached draft, use instead:
  % \node[text=bnA,fill=bnB!12,inner sep=1.5pt]
  %   at (4.375,1.56) {$\eta<2b(1-b)-2h$};

  \node[text=bnB!65!black,fill=white,inner sep=1.6pt]
    at (4.375,1.90) {$I_b=[bN,(1-b)N]$};

  \foreach \xx/\lab in {\bnXzero/0,\bnXb/bN,\bnXa/aN,
                          \bnXac/{(1-a)N},\bnXbc/{(1-b)N},\bnXend/N}{
    \draw[bnMuted,line width=0.45pt]
      (\xx,\bnYzero)--(\xx,1.03);
    \node[anchor=north,inner sep=1.5pt] at (\xx,0.99) {$\lab$};
  }

  % Phase decomposition, with cuts aligned to the graph.
  \fill[bnA!16] (\bnXzero,0.16) rectangle (\bnXa,0.49);
  \fill[bnB!28] (\bnXa,0.16) rectangle (\bnXac,0.49);
  \fill[bnC!16] (\bnXac,0.16) rectangle (\bnXend,0.49);
  \node at ({(\bnXzero+\bnXa)/2},0.325) {$A$};
  \node at ({(\bnXa+\bnXac)/2},0.325) {$B$};
  \node at ({(\bnXac+\bnXend)/2},0.325) {$C$};
  \end{tikzpicture}%
  % END TIKZ
  }
  \endgroup
  \caption{%
    (a) Schematic bottleneck geometry.
    (b) Curie--Weiss shell energies $e_k=2k(N-k)/N$.
    The shaded interval $I_b=[bN,(1-b)N]$ contains the bottleneck
    sector $B$. Dashed vertical lines mark the cuts at $aN$ and
    $(1-a)N$ with $0<b<a<1/2$.
  }
  \label{fig:bottleneck-cw}
\end{figure}

Bottleneck theory is a standard framework for proving lower bounds on the mixing time. It identifies an unavoidable intermediate regime with rare probability flow as a bottleneck to reaching the final equilibrium; see Fig. \ref{fig:bottleneck-cw} for an illustration of the bottleneck geometry. For classical Markov chain sampling, such as Glauber dynamics, it is a textbook method for proving slow-mixing results for various models \cite{LevinPeres2017}. It has been recently generalized to quantum Gibbs sampler by Gamarnik, Kiani and Zlokapa (GKZ) \cite{GKZ2024}. A convenient assumption for relating the bottleneck to mixing-time properties is that the sampling dynamics are spatially local, so that they cannot take sector $A$ into sector $C$ in one step if the sectors are chosen to be spatially distinct. While this is easily achieved in classical sampling, the quantum Gibbs sampler requires the Hamiltonian to be spatially local to apply the Lieb-Robinson bound to the Hamiltonian evolution \cite{GKZ2024}. A similar condition is assumed for quantum channels in \cite{RPBK2026}.

In this work, we consider sampling the spatially nonlocal quantum Curie--Weiss (CW) model. To address the global support of the Hamiltonian evolution, we use the spectral sectors of the CW model and a dynamical quantum bottleneck theory. In this way, we show that the sampler is (quasi-)local with respect to the spectral sectors, resulting in a mixing time that scales exponentially with the system size. This resolves the open problem of proving slow mixing for the quantum CW model in \cite{GKZ2024}. Furthermore, we find that the geometric quantum bottleneck theory proposed in \cite{GKZ2024} has a more general scope with regard to spatial locality --- it is equivalent to our dynamical quantum bottleneck theory with an appropriate bottleneck geometry. Moreover, we give a complementary interpretation of slow mixing through a diffusive process in the mean-field Fokker-Planck equation.

\section{Model, sampler, and slow-mixing result}

The quantum CW model is given by the Hamiltonian $H = H_0 + V$ with
\begin{align}\label{eq:H}
    H_0 = -\frac{1}{N} \sum_{1\leq i<j\leq N} Z_i Z_j,\quad  
    V=- h \sum_{i=1}^N X_i
\end{align}
where $X_i,Z_i$ are single-qubit Pauli operators on site $i$.
At inverse temperature $\beta>0$, the Gibbs state is given by $\rho_\beta = Z^{-1}\exp(-\beta H)$ with $Z = \tr(\exp(-\beta H))$. At low temperature $\beta\gg 1$ and weak transverse field $0<h\ll 1$, the equilibrium Gibbs state exhibits ferromagnetic order \cite{ChayesCrawfordIoffeLevit2008}.

A general framework for constructing quantum Gibbs samplers is given by Chen–Kastoryano–Gily\'en (CKG) \cite{ckg}; see also \cite{Ding-lin-lin}. Let $\{A_a\}_{a=1}^{n}$ be a family of $n$ seed jump operators, and assume $\{A_a\}^\dagger = \{A_a\}$, $\abs{\mathrm{supp}~ A_a}, \norm{A_a}<\infty$, and $n= O(N)$. The CKG Lindbladian is given by
\begin{align}\label{eq:CKG L}
    \mathcal{L}_{\mathrm{CKG}}(\rho) = -\i [B,\rho] + \sum_{a=1}^n\int_{\mathbb{R}} \mathcal{D}_{J_a(\omega)}(\rho)\ud \omega,
\end{align}
where $\mathcal{D}_L(\sigma)\equiv L \sigma L^\dagger - \frac{1}{2}\{L^\dagger L,\sigma\}$, and
\begin{align}
    J_a(\omega) := \int_{\mathbb{R}} c_\beta(\omega,t)A_a(t)\ud t,\quad A_a(t):= e^{\i H t} A_a e^{-\i H t}
\end{align}
The details of the coefficient $c_\beta(\omega,t)$ and the coherent term $B$ are summarized in App.~\ref{app: full proof}. The Lindbladian satisfies
\begin{align}
    \mathcal{L}_{\mathrm{CKG}}(\rho_\beta) = 0.
\end{align}

Let $T_t$ be a completely positive, trace-preserving (CPTP) semigroup whose infinitesimal generator is the CKG Lindbladian. The mixing time is defined as follows.
\begin{defn}
    For $0<\epsilon<1$,
    \begin{align}
        t_{\mathrm{mix},\epsilon}:= \min\{t>0: \norm{T_t(\sigma)-\rho_\beta}_1<\epsilon,\forall \sigma\}.
    \end{align}
    In particular, $t_{\mathrm{mix}}:= t_{\mathrm{mix},1/4}$.
\end{defn}

Our main application is the following lower bound for
the CKG sampler just defined.

\begin{thm}\label{thm:slow mixing main}
For fixed $\beta\gg 1$ and $0<h\ll 1$, there exists $N_0$ such that 
\begin{align}
    t_{\mathrm{mix}}(\mathcal{L}_{\mathrm{CKG}})\geq \exp( \Omega( N)), \quad N\geq N_0
\end{align}
\end{thm}
\noindent The proof combines a dynamical bottleneck theorem (Sec.~\ref{sec:bottleneck}) with equilibrium suppression of both low-energy states and local operators (Sec.~\ref{sec:suppression}).

\section{A dynamical quantum bottleneck}\label{sec:bottleneck}

Consider a generic Lindbladian $\mathcal{L}$ satisfying $\mathcal{L}(\rho) = 0$ with an exact and unique (faithful) stationary state $\rho$.
Following \cite{GKZ2024}, we take an initial state $\sigma_0:= \rho^{1/2} P_A \rho^{1/2}/p_A$ where $P_A$ is an orthogonal projector in the Hilbert space $\mathcal{H} = (\mathbb{C}^2)^{\otimes N}$ and $p_A:= \tr(\rho P_A)$. Bounding the mixing time requires estimating the overlap of $\sigma_t:= T_t(\sigma_0)$ with $P_{A^c}$, where $A^c$ is the complementary sector to $A$. We define the initial overlap
\begin{align}
    q_{A^c}:= \tr(P_{A^c} \sigma_0).
\end{align}
We also define the edge flow
\begin{align}
    J_{A^c}:= \norm{\rho^{1/2} \mathcal{L}^\dagger(P_{A^c}) \rho^{1/2}}_1,
\end{align}
and we have the following theorem.
\begin{thm}\label{thm: bottleneck}
    Let $P_{A}$ be an orthogonal projector on $\mathcal{H}$, and let $P_{A^c} = I - P_A$. Denote $p_i = \tr(P_i \rho)$ for $i = \{A,A^c\}$.
    Then the mixing time for a fixed $\epsilon\leq 2(p_{A^c} - q_{A^c}) $ is bounded from below by
    \begin{align}\label{eq:mixing time bound}
        t_{\mathrm{mix},\epsilon}\geq \frac{p_A}{J_{A^c}}\left(p_{A^c} - q_{A^c} - \epsilon/2\right)
    \end{align}
    In particular, if $p_A,p_{A^c}=\Omega(1), \; q_{A^c}=o(1),\; J_{A^c} = \exp(-\Omega(N))$, then 
    \begin{align}
        t_{\mathrm{mix}} \geq \exp(\Omega(N)).
    \end{align}
\end{thm}
\begin{proof}
    Since $0\leq P_A\leq I$, congruence by a positive operator $\rho^{1/2}$ gives $0\leq \rho^{1/2} P_A \rho^{1/2} \leq \rho$, and CPTP maps preserve the L\"owner order, giving
    \begin{align}
        0\leq \sigma_t \leq \frac{\rho}{p_A},
    \end{align}
    and a second congruence by $\rho^{-1/2}$ gives $0\leq Y_t:=p_A\rho^{-1/2} \sigma_t \rho^{-1/2} \leq I$. This gives 
    \begin{align}
        \frac{\ud}{\ud t}\tr(P_{A^c} \sigma_t) &= \tr(P_{A^c} \mathcal{L}(\sigma_t)) \nonumber\\
        &= p_A^{-1}\tr( Y_t \rho^{1/2} \mathcal{L}^\dagger(P_{A^c}) \rho^{1/2}) \nonumber\\
        &\leq  p_A^{-1} J_{A^c}.
    \end{align}
 Integrating over $t$ gives
    \begin{align}
        \tr(P_{A^c} \sigma_t) \leq  q_{A^c} + t p_A^{-1} J_{A^c}
    \end{align}
    Applying Helstrom's theorem $\norm{\sigma_t - \rho}_1 \geq 2 \abs{\tr(P_{A^c}(\sigma_t -\rho))}$ and the definition of the mixing time, we arrive at \eqref{eq:mixing time bound}.
\end{proof}

Theorem~\ref{thm: bottleneck} reduces to the classical bottleneck theory in App.~\ref{app: classical} for classical and discrete dynamics. By choosing the transition rate $r(x,y)\geq 0$ for configurations $x\neq y$ in the computational basis, we have
\begin{align}
    \LL^\dagger(P_{A^c}) = \frac{1}{2}\sum_{x\in A}r(x,A^c)|x\rangle\langle x| - \frac{1}{2}\sum_{x\in A^c}r(x,A)|x\rangle\langle x|.
\end{align}
Consequently, 
\begin{align}
    J_{A^c}  =  \sum_{x\in A} \pi(x) r(x,A^c),
\end{align}
where we have used stationarity of the distribution $\pi(x)$ so that $\sum_{x\in A} \pi(x) r(x,A^c) = \sum_{x\in A^c} \pi(x) r(x,A)$. Identifying the transition rate with the Markov chain transition probability at a unit time step, we find that $J_{A^c}$ and $J_{A^c}/p_A$ correspond to the edge measure and conductance defined in App.~\ref{app: classical}. However, upon including quantum coherence, Theorem~\ref{thm: bottleneck} differs from the classical bottleneck theory in two respects. First, \eqref{eq:mixing time bound} contains a quantum contribution $q_{A^c}$ due to noncommutativity, which is absent in the classical limit.
Second, the edge flow $J_{A^c}$ depends not only on the stationary distribution within subspace $A$ or $A^c$, but also on the full equilibrium density matrix.

\section{Equilibrium suppression}\label{sec:suppression}

It is convenient to rewrite the Hamiltonian \eqref{eq:H} (up to a constant shift) using the collective variables:
\begin{align}
    H_0 = \frac{2 K (N-K)}{N},\quad V  = - h  M_x,
\end{align}
where $M_x := \sum X_i$, $M_z := \sum Z_i$ and $K := (N-M_z)/2$. In the computational basis, $K$ is the number of downward spins. Let $P_k$ denote the orthogonal projection onto the eigenspace of $H_0$ with energy $e_k  = 2k(N-k)/N$, and call the subspace $P_k$ the $k$-shell. The transverse field term changes $K$ by one: $P_k M_x P_l = 0, \abs{k-l}\neq 1$. Therefore, the spectrum of $H_0$ forms a one-dimensional $N+1$-site graph, $\{0,1\ldots, N\}$, 
and each action of the transverse field term performs a ``quantum walk'' between nearest-neighbor shells with a potential given by $e_k$.

% On this one-dimensional spectrum graph, define the distance between different shells $k,l$ and the subset of shells $S,T$ as
% \begin{align}
%     d(k,l) = \abs{k-l},\quad d(S,T) = \min_{k\in S, l \in T}\abs{k-l}
% \end{align}

Let us parameterize the spectrum graph as follows. Take $0<b<a<1/2$, and define the forbidden interval as
\begin{align}
    I_b:= \{ bN, \ldots, (1-b)N \}
\end{align}
where $bN$ and $(1-b)N$ are understood as nearby integers (the rounding error can be ignored for sufficiently large $N$). For $k\in I_b$, define its depth within the interval as  
\begin{align}
    d_{I_b}(k):= \min\{k-bN +1 , (1-b)N  - k + 1\}
\end{align}
The bottleneck geometry is set by
\begin{align}
    &P_A:= P_{0\leq k \leq aN},\quad  P_C := P_{(1-a)N\leq k \leq N} \nonumber\\
    &P_B:= I - P_A - P_C
\end{align}
Hence, every shell of $B$ and the cuts $\partial A$ and $\partial C$ are inside $I_b$ and have depth $d_{I_b}\gtrsim (a-b)N $. The parameterization is visualized in Fig.~\ref{fig:bottleneck-cw}.

We begin by establishing that the low-energy subspaces have small weight in high-energy shells; this is reminiscent
of Theorem 2.2 in \cite{Arad}. In particular, let $E_0$ be the ground-state energy of $H$, and let $\Pi_{(-\infty, E]}(H)$ be an orthogonal projection onto the subspace spanned by eigenstates of $H$ with energies below $E$. \begin{prop}\label{prop:1 main}
    Assume $2b(1-b)>2h+\eta$. For any $k\in I_b$, 
    \begin{align}
        \norm{P_k \Pi_{(-\infty, E_0+\eta N]}(H)} \leq C e^{-c d_{I_b}(k)},
    \end{align}
    where $C,c>0$ are constants independent of $N$.
\end{prop}
\noindent This estimate indicates that the low-energy subspace of $H_0$ would be a good approximation to that of $H$. In particular, if we insert $\Pi+(I-\Pi)$ into $p_B:= \tr(P_B \rho_\beta)$, the term $\tr(P_B \Pi \rho_\beta)$ would be exponentially suppressed by Proposition~\ref{prop:1 main}.
Observing that the thermal weights are exponentially suppressed above the energy $E_0+\eta N$ with $\beta \eta>\log 2$, we find that the term $\tr(P_B (I-\Pi) \rho_\beta)$ is also exponentially suppressed. Together, they imply $p_B\leq \exp(-\Omega(N))$. Using the symmetry operator $U = \prod_i X_i$ and $U\rho_\beta U =\rho_\beta,\; U P_A U = P_C$, we have $p_A =p_C = (1-p_B)/2 =\Omega(1)$. Moreover, the initial overlap $q_{C}:= \tr(P_{C} \sigma_0)$ can be shown to be exponentially suppressed by commuting $P_A$ across the Gibbs state and noting that $[H,P_A]$ is located near $\partial A$ and thus inside $I_b$.

We now turn to the dynamics. Since both the dissipative and coherent parts of the CKG sampler \eqref{eq:CKG L} contain Gaussian filter functions in their time integrals, the Hamiltonian evolution can be truncated.
\begin{prop}
Let $\mathcal{L}_{\mathrm{CKG},\Gamma}$ denote the CKG Lindbladian with the time integral truncated to $\abs{t}\leq \beta \Gamma$.
    Suppose $\norm{A_a}\leq A_0$. Then
    \begin{align}\label{eq: diamond 0}
        \norm{\mathcal{L}_{\mathrm{CKG}}-\mathcal{L}_{\mathrm{CKG},\Gamma}}_{\diamond} \leq 5 n A_0^2(e^{-\Gamma^2} + e^{-\Gamma})
    \end{align}
\end{prop}
\noindent Consequently, by choosing $\Gamma = N$, we can replace the full CKG Lindbladian with the truncated one up to an exponentially small error. The remaining task is to estimate the edge flow for these jump operators under truncated Heisenberg time evolution. Although the jump operator grows into a global operator under finite Heisenberg time evolution because of the lack of geometric locality, it can inject only finite energy because the local interaction strength is finite \cite{Arad}:
\begin{align}
    \Lambda:= \max_i \sum_{X \ni i}\norm{h_X} = \frac{N-1}{N}+h<\infty,
\end{align}
where we write $H = \sum_X h_X$. Indeed, we have the following lemma.
\begin{lem}\label{lem: seed}
    Consider an operator of the form $\mathcal{O}=S_q(t_q)\cdots S_1(t_1)$ with finite $q$. When $\abs{t_j}\leq \beta N$,
    \begin{align}
        \norm{[P_A,\mathcal{O}]\rho_\beta^{1/2}}_2 \leq   e^{-\Omega(N)}.
    \end{align}
    The same holds for $P_C$.
\end{lem}
\noindent To show Lemma \ref{lem: seed}, we similarly insert the low-energy subspace projectors between the commutator and the Gibbs state. Since there are at most $3q$ operators in $\mathcal{O}$ and a finite number of operators inject finite energy, this product cannot bring states near $\partial A$ into states that are beyond $I_b$ and have low energy, resulting in the exponential suppression. By taking $\mathcal{O} = A_a(t)$ and using Proposition \ref{prop:1 main}, we can show that
\begin{align}
    J_C\leq  e^{-\Omega(N)}
\end{align}

Combining the above results with Theorem~\ref{thm: bottleneck}, we arrive at Theorem~\ref{thm:slow mixing main}; see App.~\ref{app: full proof} for a full proof.

\section{Connection with GKZ}

In many situations, including the quantum CW model, we encounter the bottleneck geometry in Fig.~\ref{fig:bottleneck-cw}. If the sampling algorithm is taken to be local with respect to the bottleneck geometry, i.e., a one-step update cannot connect sectors $A$ and $C$, then Theorem~\ref{thm: bottleneck} can be transformed into the geometric bottleneck theory in GKZ \cite{GKZ2024}; see the classical counterpart in App.~\ref{app: classical}. However, the actual quantum Gibbs sampler is always approximately local due to the Hamiltonian evolution. To estimate the error of the approximate dynamics, GKZ assumes spatial locality of the Lindbladian and uses the Lieb-Robinson bound. As already suggested by GKZ, the appropriate distance for defining locality need not be that between spatial lattice sites; it could instead be a distance in the spectrum. Indeed, our previous discussion reveals that sampling the quantum CW model is approximately local on the spectrum graph, although it has global support on the spatial lattice. The remaining question is whether there exists a nearby Lindbladian that is exactly local on the spectrum graph. This is achieved by the following proposition.
% CITATION-SCOPE: GKZ2024 Sections 3.1--3.2 already define spectral/operator-locality distances; spatial locality enters specific applications rather than the general bottleneck theorem.
\begin{prop}\label{thm: GKZ}
    Let $P_A,P_C$ be orthogonal projectors in $\mathcal{H}$, and let $P_B = 1 - P_A - P_C$. Denote $p_i = \tr(P_i \rho)$ for $i = \{A,B,C\}$. Assume 
    \begin{align}
        p_A,p_C = \Omega(1),\quad  p_B, q_C \leq e^{-\Omega(N)}.
    \end{align}
    If there exists a Lindbladian $\tilde{ \mathcal{L}}$ such that
    \begin{align}\label{eq: nearby L condition}
        &\tilde{\mathcal{L}}^\dagger(P_C) = (1-P_A) \tilde{\mathcal{L}}^\dagger(P_C)(1-P_A) \nonumber \\ &\norm{\mathcal{L} - \tilde{\mathcal{L}}}_{\diamond} \leq e^{-\Omega(N)}
    \end{align}
    then
    \begin{align}
            J_C \leq e^{-\Omega(N)}.
    \end{align}
    Conversely, if $J_C$ is exponentially small, then there exist subprojections $P_A'\leq P_A,\; P_C'\leq P_C$ and a Lindbladian $\tilde{ \mathcal{L}}$ such that
    \begin{align}
        p_A',p_C' = \Omega(1),\quad p_B', q_C' \leq e^{-\Omega( N)} 
    \end{align}
    and \eqref{eq: nearby L condition} is satisfied with respect to the subprojections.
\end{prop}
\noindent The proof of the forward direction is straightforward. The proof of the converse direction uses the idea of quantum hypothesis testing to find the subprojections $P'_{A}$ ($P'_{C}$) by subtracting the part most affected by sector $C$ ($A$); see App.~\ref{app: GKZ}.

Therefore, we have established a formal connection between the dynamical bottleneck theory in Theorem \ref{thm: bottleneck} and the geometric bottleneck theory from GKZ.

\section{Mean-field analysis}

In this section, we aim to study the large-$N$ limit of the Lindbladian dynamics and derive an effective Fokker-Planck equation. This approach has been applied to the classical CW model under Glauber dynamics in \cite{gwl}, and it was shown that the relaxation time for the magnetization distribution is exponentially large in $N$. We will see that a similar effective Fokker-Planck equation emerges for the quantum model.

Let us begin with the equilibrium distribution in the infinite-$N$ limit. For the product-state ansatz $\rho_{\mathrm{MF}}\propto (e^{-\beta H_{\mathrm{MF}}})^{\otimes N}$ with
\begin{align}
    H_{\mathrm{MF}} = - m Z-h X,
\end{align}
the two wells in the ferromagnetically ordered phase are located at $m_{\pm} = \pm \sqrt{m_\beta^2 - h^2}$
with $m_\beta$ given by the solution of $m_\beta = \tanh(\beta m_\beta)$. The saddle point, however, has $m = 0$ and is described by the product state of $(I+x_s X)/2$ with $x_s:=\tanh(\beta h)$. 

Now, applying the CKG Lindbladian with $H_{\mathrm{MF}}$ to a general single-qubit state $\rho(t) = (I+x(t)X+y(t)Y+z(t) Z)/2$ and expanding near the saddle point $(x,y,z)=(x_s,0,0)$ to leading order in $N$, we find (App.~\ref{app: mean field})
\begin{align}\label{eq: eom saddle}
    \delta\dot{ x} = -\gamma_x \delta x,\quad \delta\dot{ y} = -\gamma_y \delta y,\quad \delta\dot{ z} = \gamma_z \delta z
\end{align}
with $\gamma_{x},\gamma_{y},\gamma_{z}>0$. Thus, the saddle point is unstable in the $z$-direction under the CKG Lindbladian.

To capture the fluctuations near the saddle point, we need to keep the leading $1/N$-contributions. With a slight abuse of notation, let us denote the magnetization density by $m=N^{-1}\sum_i Z_i$, which has zero expectation value at the saddle point. For an arbitrary smooth function $f(m)$, we find (App.~\ref{app: mean field})
\begin{align}\label{eq: operator FP}
    \mathcal{L}^\dagger f( m) \approx \gamma_z m f'( m) + \frac{D}{N} f''( m) + O(N^{-2})
\end{align}
with $D:= \frac{N}{2} \langle \Gamma(m) \rangle_{\mathrm{MF}}$, where \begin{align}
    \Gamma(\mathcal{O}):= \sum_a \int \ud\omega~ [J_a(\omega),\mathcal{O}]^\dagger[J_a(\omega),\mathcal{O}] 
\end{align}
is the carr\'e du champ operator \cite{Lindblad1976,Meyer1976CarreDuChamp}. Since \eqref{eq: operator FP} is diagonal in the eigenbasis of $m$, we can assign a probability distribution $p(m,t)$ to it, yielding the following Fokker-Planck equation:
\begin{align}\label{eq:FP}
    \partial_t p = - \partial_m(\gamma_z  m p)+ \frac{D}{N} \partial_m^2 p 
\end{align}
The equilibrium distribution of this Fokker-Planck equation is given by
\begin{align}
    p_{\mathrm{eq}}(m) = p_{\mathrm{eq}}(0)\exp\left(\frac{N\gamma_z}{2 D}m^2 \right)
\end{align}
Writing $p(m,t)=q(m,t)p_{\mathrm{eq}}(m)$, we find
\begin{align}
    \partial_t q = \gamma_z m \partial_m q + \frac{D}{N}\partial_m^2 q
\end{align}
The static profile of $q=q(m)$ is given by
\begin{align}\label{eq:qm}
    q(m)=\frac{q_+ + q_-}{2}+ \frac{q_+ - q_-}{2}\mathrm{erf} \left( \sqrt{\frac{N\gamma_z}{2 D}} m\right),
\end{align}
with $q_{\pm} = q(m=\pm \infty)$ specifying the boundary conditions. The Fokker-Planck equation \eqref{eq:FP} can be regarded as a current-conservation equation, with the probability current given by
\begin{align}
    j(m):= \gamma_z m p - \frac{D}{N}\partial_m p = - \frac{D}{N}p_{\mathrm{eq}} \partial_m q.
\end{align}
At sufficiently large $N$, $q(m)$ in \eqref{eq:qm} is nearly a step function near $m=0$, indicating that $j(m)$ is concentrated at $m=0$. Since the magnetization distribution at the saddle point is exponentially suppressed, $p_{\mathrm{eq}}(0)\sim \exp(-\Omega(N))$ (see \cite{ChayesCrawfordIoffeLevit2008,gwl}), the probability current $j(0)\propto p_{\mathrm{eq}}(0)$ is also exponentially suppressed. Therefore, the relaxation time for the distribution to reach equilibrium, i.e., $q(m,t)\to 1$, is exponentially large in $N$.
% CITATION-SCOPE: These sources support the quantum phase diagram and classical metastability, not by themselves the exact scalar saddle-density/current-concentration claims.

\section{Discussion}

\emph{i).} Our mixing-time results can be generalized to more general quantum perturbations of the classical CW model beyond the transverse field. Specifically, we allow for $k$-local perturbations that involve at most $k$-body interactions, because they still inject only finite energy and ``walk'' within a finite neighborhood along the spectrum graph. Moreover, some implicit assumptions for the quantum CW model are unnecessary for proving the mixing-time bound: first, the global $\mathbb{Z}_2$ symmetry can be broken as long as $p_A,p_C= \Omega(1)$; second, if we break the detailed balance condition by including the Hamiltonian itself in the coherent term of the Lindbladian, the conclusion of slow mixing remains valid.

\emph{ii).} Let us contrast the CKG Lindbladian with the Davies generator \cite{Davies1974}. The key difference is that the Davies generator resolves every Bohr frequency however small it is; in particular, it resolves the two wells even though their energy splitting is exponentially small. This resolution allows for finite coherent tunneling between the two wells, violating the assumptions of the bottleneck theorem \ref{thm: bottleneck}. In particular, let us assume that the degrees of freedom are just the two GHZ states associated with the two wells.
Consider the jump operator $Z_i$ that acts on the GHZ state as $Z_i \approx m_i \tau_x$, where $m_i$ is some effective coefficient and $\tau_x$ flips between the two GHZ states. Let $\Pi_E$ be the spectral projection of $H$, and $Z_{i,\omega}:= \sum_{E'-E = \omega} \Pi_{E'} Z_i \Pi_E$.
Then, $Z_{i,\delta} \approx  m_i \tau_+$ and $Z_{i,-\delta} \approx m_i \tau_-$ with $\delta \sim e^{-N}$, and the Davies generator is approximated by $\LL_{\mathrm{D}} \sim \mathcal{D}_{\tau_+} + \mathcal{D}_{\tau_-}$. Since $\LL^\dagger_{\mathrm{D}}(\tau_x) \approx  -\tau_x$, the magnetization relaxes at a finite rate. In contrast, the CKG Lindbladian cannot resolve energy splittings much smaller than $1/\beta$. Specifically, given $Z_i(t)\approx m_i(e^{\i \delta t} \tau_+ + e^{-\i \delta t}\tau_-)$, we have  $J_i(\omega)\approx m_i c_\beta(\omega)(\tau_+ + \tau_-)$, so that $\LL_{\mathrm{CKG}} \propto \mathcal{D}_{\tau_+ + \tau_-} = \mathcal{D}_{\tau_x}$, which does not relax magnetization. Since the slow modes are precisely unstable to coherent tunneling, we expect that the Davies generator would have at most polynomial-in-$N$ mixing time.

To rigorously show that the Davies generator has a parametrically faster mixing time, one needs to establish an upper bound on the mixing time using the spectral gap, which is beyond the scope of this work. Furthermore, an interesting question is whether there is always a parametric separation of mixing times between GNS-detailed-balance (e.g. the Davies generator) and KMS-detailed-balance (e.g. the CKG sampler) Lindbladians in ``ordered'' quantum states.

\emph{iii).} Our quantum bottleneck theory can be applied to the preparation of other nonlocal Gibbs states using the CKG Lindbladian. For example, preparing the quantum low-density-parity-check (LDPC) code would require a geometrically nonlocal CKG Lindbladian. Due to the linear confinement and linear distance properties \cite{PanteleevKalachev2022Good, HongGuoLucas2025}, we expect slow mixing of the Lindbladian. Furthermore, as the classical LDPC code under quantum perturbations realizes nontrivial ``orders'' \cite{YinNandkishoreLucas2024Eigenstate}, we expect slow mixing of the CKG Lindbladian in that case as well.

\emph{iv).} Slow mixing of a Lindbladian is often related to metastability. Classically, a small conductance implies a metastable phase, which provides a source for Hamiltonian learning tasks \cite{JayakumarEtAl2026Metastable}. Metastability has been generalized to quantum Lindbladians \cite{MacieszczakEtAl2016Metastability,BergamaschiChenVazirani2025Metastability}; therefore, we expect that our edge flow, which is essentially the quantum conductance, would be a useful metric for quantum metastable phases and further applications.

\emph{Acknowledgments.}--- OpenAI GPT 6 was used to prototype the results and help prove the converse direction of Proposition~\ref{thm: GKZ}, and all the outputs were verified by the authors. Research at Perimeter Institute is supported in part by the Government of Canada through the Department of Innovation, Science and Economic Development and by the Province of Ontario through the Ministry of Colleges and Universities. 

\bibliographystyle{apsrev4-1}
\bibliography{references}
\newpage
\onecolumngrid

\appendix
\renewcommand{\thesubsection}{\thesection.\arabic{subsection}}

\section{A brief review of classical bottleneck theory}\label{app: classical}

Following \cite{LevinPeres2017}, a Markov chain is described by a sequence of random variables $(X_0,X_1,\ldots)$ taking values in the configuration space $\mathcal{X}$ and a transition matrix $P$. The transition matrix determines the next step from the current state $x\in \mathcal{X}$, and we denote the transition probability by $P(x,\cdot)$. A stationary state $\pi$ satisfies $\pi(y) = \sum_{x\in \mathcal{X}} \pi(x) P(x,y)$. We assume that the Markov chain is irreducible and aperiodic so that there exists a unique stationary state $\pi$.

The edge measure $Q(x,y)$ is defined as
\begin{align}
    Q(x,y):= \pi(x)P(x,y),\quad Q(A,B):= \sum_{x\in A, y\in B} \pi(x)P(x,y)
\end{align}
For a stationary chain $\{X_t\}$ that starts in the set $A$, the probability of reaching its complement $A^c$ is bounded by
\begin{align}
    \mathbb{P}\{X_0\in A, X_t\in A^c\}\leq \sum_{i=1}^t\mathbb{P}\{X_{i-1}\in A, X_i\in A^c\} = t Q(A,A^c)
\end{align}
Dividing both sides by $\pi(A)$, we find that there exists a state $x\in A$ such that the transition probability to the complement after $t$ steps is at most
\begin{align}\label{eq: app classical conductance}
    P^t(x,A^c) \leq t \frac{Q(A,A^c)}{\pi(A)}=:t \Phi(A),
\end{align}
where $\Phi(A)$ is called the conductance of the set $A$. For a classical Markov chain, the mixing time is the minimum $t$ such that 
\begin{align}
    \max_{x\in \mathcal{X}}\norm{P^t(x,\cdot) - \pi}_1\leq \epsilon
\end{align}
holds. Using $2\abs{P^t(x,A^c) - \pi(A^c)}\leq \norm{P^t(x,\cdot) - \pi}_1$ and \eqref{eq: app classical conductance}, we find
\begin{align}\label{eq: app mixing time conductance}
    t_{\mathrm{mix}}\geq \Phi(A)^{-1} (\pi(A^c) - \epsilon/2),
\end{align}
where we have assumed $\pi(A^c) \geq \epsilon/2$.

Now suppose that the total configuration space is partitioned into sets $A,B,C$ such that $P(A,C) = 0$, meaning that a one-step transition from the set $A$ to the set $C$ is forbidden. Then the conductance of the set $A$ satisfies
\begin{align}
    \Phi(A) = \frac{Q(A,A^c)}{\pi(A)} = \frac{\sum_{x\in A,y\in B} \pi(x) P (x,y)}{\pi (A)}\leq \frac{\sum_{y\in B} \sum_{x\in\mathcal{X}} \pi(x) P (x,y)}{\pi (A)} = \frac{\pi(B)}{\pi(A)},
\end{align}
where we used $P(x,y)\geq 0$. Substituting this bound into \eqref{eq: app mixing time conductance} gives
\begin{align}
    t_{\mathrm{mix}}\geq \frac{\pi(A)}{\pi(B)}(\pi(A^c) - \epsilon/2).
\end{align}
In particular, if $\pi(A),\pi(C) = \Omega(1)$ and $\pi(B) = \exp(-\Omega(N))$, then $t_{\mathrm{mix}}\geq \exp(\Omega(N))$.

\section{Proof of the connection with GKZ}\label{app: GKZ}

In this section, we give a full proof of Proposition \ref{thm: GKZ}. We begin with the forward direction.

\begin{proof}[Proof of the forward direction of Proposition \ref{thm: GKZ}]
     First, we can approximate the edge flow using $\tilde \LL^\dagger$ because
    \begin{align}
        \norm{\rho^{1/2}(\LL^\dagger - \tilde \LL^\dagger)(P_C) \rho^{1/2}}_1\leq \norm{(\LL^\dagger - \tilde \LL^\dagger)(P_C)} \leq e^{-\Omega(N)}
    \end{align}
    Next, we write
    \begin{align}
        X:=\tilde \LL^\dagger(P_C) = \sum_{i,j=B,C}P_i \tilde\LL^\dagger(P_C)P_j=: P_C X P_C + Y
    \end{align}
    Using the GKSL form of the Lindbladian, one can show that $P_CX P_C \leq 0$, which yields
    \begin{align}
        \norm{\rho^{1/2}P_C\tilde \LL^\dagger(P_C)P_C \rho^{1/2}}_1 = - \tr(\rho P_C XP_C)  = -\tr(\rho  X)+\tr(\rho Y) \leq \abs{\tr(P_C (\tilde \LL - \LL)\rho)} + \norm{\rho^{1/2}Y \rho^{1/2}}_1
    \end{align}
    Hence, 
    \begin{align}
        \norm{\rho^{1/2} \tilde \LL^\dagger(P_C) \rho^{1/2}}_1 &=\sum_{i,j=B,C} \norm{\rho^{1/2}P_i\tilde \LL^\dagger(P_C)P_j \rho^{1/2}}_1 \nonumber\\
        &\leq \abs{\tr(P_C (\tilde \LL - \LL)\rho)} + 2\norm{\rho^{1/2}Y \rho^{1/2}}_1 \nonumber\\
        &\leq\norm{\tilde \LL - \LL}_\diamond  +2 \norm{\tilde \LL^\dagger(P_C)}(p_B + 2 \sqrt{p_B p_C})\nonumber\\
        &\leq e^{-\Omega( N)}
    \end{align}
    where we used $\norm{\tilde \LL^\dagger(\mathcal{O})}\leq \mathrm{poly}(N)$ for a bounded operator $\mathcal{O}$. Together, these estimates complete the proof.
\end{proof}

To prove the converse direction, consider the GKSL form of the Lindbladian
\begin{align}
    \mathcal{L}(\rho)=-i[B, \rho]+\sum_j J_j \rho J_j^{\dagger}-\frac{1}{2}\{G, \rho\}, \quad G=\sum_j J_j^{\dagger} J_j, \quad B=B^{\dagger} .
\end{align}
with $\norm{G},\norm{B}\leq 1$.
Let us denote
\begin{align}
    R:= P_A \mathcal{L}^\dagger(P_C) P_A,\quad T:= P_C \mathcal{L}^\dagger(P_C) P_A
\end{align}
Define the variables
\begin{align}
    \lambda:= \norm{P_C \sqrt{\rho}P_A}_2^2,\quad \ell := p_B+\lambda,\quad \epsilon:= J_C+4\sqrt{2}\sqrt{\ell}
\end{align}
and 
\begin{align}
    \Phi(S):= \sum_j J_j^\dagger S J_j. 
\end{align}
\begin{lem}\label{lem: GKZ 1}
    \begin{align}
        \tr(\rho R)\leq \epsilon,\quad \norm{\sqrt{\rho} T\sqrt{\rho}}_1 \leq \epsilon
    \end{align}
\end{lem}
\begin{proof}
    Observe that
    \begin{align}
        &\norm{(1-P_A)\sqrt{\rho}P_A}_2^2 \leq \norm{P_B\sqrt{\rho}P_A}_2^2 + \norm{P_C\sqrt{\rho}P_A}_2^2 \leq p_B + \lambda =: \ell\nonumber\\
        &\norm{(1-P_C)\sqrt{\rho}P_C}_2^2 \leq \norm{P_B\sqrt{\rho}P_C}_2^2 + \norm{P_A\sqrt{\rho}P_C}_2^2 \leq p_B + \lambda =: \ell
    \end{align}
    so that $\norm{[\sqrt{\rho},P_A]}_2\leq\sqrt{2\ell}$ and $\norm{[\sqrt{\rho},P_C]}_2\leq\sqrt{2\ell}$.
    Then
    \begin{align}
        \norm{P_A \rho P_A - \sqrt{\rho} P_A \sqrt{\rho}}_1\leq (\norm{P_A \sqrt{\rho}}_2 + \norm{\sqrt{\rho} P_A }_2 )\norm{P_A \sqrt{\rho} - \sqrt{\rho} P_A}_2 = 2 \sqrt{2 p_A \ell}
    \end{align}
    which yields
    \begin{align}
        \tr(\rho R) = \tr(P_A\sqrt{\rho}\mathcal{L}^\dagger(P_C)\sqrt{\rho}) + \tr(\mathcal{L}^\dagger(P_C)(P_A \rho P_A - \sqrt{\rho} P_A \sqrt{\rho})) \leq J_C + 4\sqrt{2\ell} =:\epsilon,
    \end{align}
    where we used $p_A\leq 1$. Next,
    \begin{align}
        \norm{\sqrt{\rho}T\sqrt{\rho}}_1 &\leq  \norm{\sqrt{\rho}T\sqrt{\rho} - P_C \sqrt{\rho} \mathcal{L}^\dagger(P_C)\sqrt{\rho} P_A}_1 + J_C \nonumber\\
        &\leq J_C + 2\sqrt{2\ell}(\sqrt{p_A} + \sqrt{p_C})\nonumber\\
        &\leq J_C + 4 \sqrt{\ell} \leq \epsilon
    \end{align}
\end{proof}

Define two unnormalized states
\begin{align}
    \sigma = T P_A \rho P_A T^\dagger,\quad \omega = P_C \rho P_C
\end{align}
and the orthogonal projection onto the eigenstates with positive eigenvalues:
\begin{align}
    E_C:= \Pi_{(0,\infty)}(\sigma - \omega)
\end{align}
\begin{lem}\label{lem: GKZ 2}
    \begin{align}
        \tr(\rho E_C) + \tr(\sigma (P_C - E_C)) \leq \epsilon
    \end{align}
    Moreover, since $0\leq E_C\leq P_C$, we have
    \begin{align}\label{eq: GDK final 1}
        &0\leq \tr(\rho E_C)\leq \epsilon \nonumber\\
        &0\leq \tr(\sigma (P_C - E_C))\leq \epsilon
    \end{align}
\end{lem}
\begin{proof}
    Using $\tr(\rho E_C) = \tr(\omega E_C)$ and $\tr((\sigma-\omega)E_C) = (\tr(\abs{\sigma-\omega}) + \tr(\sigma-\omega))/2$, we have
    \begin{align}
        \tr(\rho E_C) + \tr(\sigma (P_C - E_C)) = \frac{\tr\sigma+\tr\omega - \norm{\sigma-\omega}_1}{2} \leq \norm{\sqrt{\sigma} \sqrt{\omega}}_1
    \end{align}
    where we used the Powers–St\o rmer inequality in the second step. Using the equivalence of the 1-norm under polar decompositions, we obtain
    \begin{align}
        \norm{\sqrt{\sigma} \sqrt{\omega}}_1 = \norm{ \sqrt{\omega} T \sqrt{\rho}}_1 = \norm{ \sqrt{\rho} T \sqrt{\rho}}_1 
    \end{align}
    Applying Lemma \ref{lem: GKZ 1} completes the proof.
\end{proof}
Denote $P_C':= P_C - E_C$ and $W:= R + T^\dagger P_C' T \geq 0$. Using Lemma \ref{lem: GKZ 1} and Lemma \ref{lem: GKZ 2}, we have
\begin{align}
    \tr(\rho W) = \tr(\rho R) + \tr(\sigma P_C')\leq 2\epsilon
\end{align}
Define, for $0<\theta<1$,
\begin{align}
    E_A: = \Pi_{(\theta,\infty)}(W)
\end{align}
and $P_A':= P_A - E_A\geq 0$. First, since $P_A' R P_A'\leq P_A' W P_A' \leq \theta P_A'$, we have
\begin{align}\label{eq: modified region bound 1}
    \norm{P_A' \mathcal{L}^\dagger(P_C')P_A'} \leq  \norm{P_A' \mathcal{L}^\dagger(P_C)P_A'}\leq \theta,
\end{align}
where we used $P_A'\mathcal{L}^\dagger(P_C')P_A' = \Phi(P_C') \leq  \Phi(P_C) = P_A'\mathcal{L}^\dagger(P_C)P_A' $. Second, 
\begin{align}
    \norm{P_A'\mathcal{L}^\dagger(P_C') P_C' } &\leq  \norm{P_A'\mathcal{L}^\dagger(P_C) P_C' } + \norm{P_A'\Phi(E_C) P_C' } \nonumber\\
    &\leq  \norm{P_C' T P_A'} + \sqrt{\norm{P_A'\Phi(E_C) P_A'}  \norm{P_C'\Phi(E_C) P_C'} } \nonumber\\
    &\leq 2\sqrt{\theta} 
\end{align}
where in the second step we used the Cauchy–Schwarz inequality, and in the third step we used $P_A' T^\dagger P_C' T P_A'\leq P_A'WP_A'\leq \theta P_A'$ and $P_A'\Phi(E_C) P_A'\leq P_A' R P_A'\leq \theta P_A'$. Third,
\begin{align}
    \tr(\rho E_A)\leq \theta^{-1} \tr(\rho W) \leq \frac{2\epsilon}{\theta}.
\end{align}
\begin{proof}[Proof of the converse direction of Proposition \ref{thm: GKZ}]
    For each jump operator $J_j$, define the modified operator
    \begin{align}
        K_j:= J_j-D_j,\quad D_j:= P_C' J_j P_A'
    \end{align}
    then $P_C' K_j P_A' = 0$. From \eqref{eq: modified region bound 1},
    \begin{align}
        \norm{\sum_j D_j^\dagger D_j} = \norm{P_A'\Phi(P_C')P_A'} = \norm{P_A' \mathcal{L}^\dagger(P_C')P_A'}\leq \theta
    \end{align}
    Applying this bound to the difference between the dissipative parts of the two Lindbladians gives
    \begin{align}
        \norm{\mathcal{D}_J -\mathcal{D}_K }_{\diamond}\leq 4 \sqrt{\theta} + 2 \theta 
    \end{align}
    For the modified coherent term $\tilde B$, we change only the part $P_A' B P_C'$ to satisfy
    \begin{align}
        P_A'\tilde B P_C' = - \frac{\i}{2}P_A'\tilde G P_C' 
    \end{align}
    so that
    \begin{align}
        P_A'\tilde{\mathcal{L}}^\dagger(P_C') = \i P_A'\tilde B P_C' - \frac{1}{2} P_A'\tilde G P_C' = 0.
    \end{align}
    The error introduced by modifying the off-diagonal coherent term is controlled by
    \begin{align}
        \norm{P_A'(\tilde B - B)P_C'} &\leq \frac{1}{2}\norm{P_A'(\tilde G - G)P_C'} + \norm{P_A'\mathcal{L}^\dagger(P_C')P_C'} + \norm{P_A' \Phi(P_C')P_C'} \nonumber\\
        &\leq \frac{1}{2}(2\sqrt{\theta}+\theta)+ 2\sqrt{\theta}+\sqrt{\theta} = 4\sqrt{\theta} + \frac{1}{2}\theta
    \end{align}
    This yields 
    \begin{align}
        \norm{\mathcal{L} - \tilde{\mathcal{L}}}_{\diamond}\leq 4\sqrt{\theta}+2\theta + 2(4\sqrt{\theta} + \frac{1}{2}\theta) = 12\sqrt{\theta}+3\theta
    \end{align}
    Choosing $\theta = \epsilon^{2/3}$, we thus obtain
    \begin{align}\label{eq: GDK final 2}
        \tr(\rho E_A)\leq 2 \epsilon^{1/3},\quad \norm{\mathcal{L} - \tilde{\mathcal{L}}}_{\diamond}\leq 15 \epsilon^{1/3}
    \end{align}
    Finally, define the new initial state $\sigma_0':= \sqrt{\rho}P_A'\sqrt{\rho}/p_A'$. We then have
    \begin{align}\label{eq: GDK final 3}
        q_C':= \tr(P_C' \sigma_0') = \frac{\norm{P_C' \sqrt{\rho}P_A'}_2^2}{p_A'}\leq \frac{\norm{P_C \sqrt{\rho}P_A}_2^2}{p_A'} = \frac{\lambda}{p_A'} = \frac{p_A }{p_A'}q_C.
    \end{align}
    Now, using the assumption in Theorem \ref{thm: GKZ}, we can choose $\epsilon\leq \mathrm{poly}(N) e^{-c N}$, which, combined with \eqref{eq: GDK final 1}, \eqref{eq: GDK final 2}, and \eqref{eq: GDK final 3}, completes the proof. 
\end{proof}

\section{The full proof of the mixing-time bound}\label{app: full proof}

\subsection{Static properties of the Gibbs state}

\begin{lem}\label{lem: 1}
    Let $\Pi_{(-\infty, E]}(H)$ be an orthogonal projection onto the eigenstate subspace of $H$ with energy lower than $E$. Suppose that, for any $k\in I = [k_-,k_+]$, 
    \begin{align}
        e_k - E \geq \Delta N - c_0,
    \end{align}
    with $c_0$ a constant and $\Delta > 2h$. Then, there exists a constant $\lambda \in(0,1)$ such that
    \begin{align}
        \norm{P_k \Pi_{(-\infty, E]}(H)}\leq\lambda ^{k-k_-+1} +  \lambda ^{k_+-k+1}.
    \end{align}
\end{lem}
\begin{proof}
   For brevity, write $R = \Pi_{(-\infty, E]}(H)$. 
    For $\psi\in \mathrm{Ran}R$, 
    \begin{align}\label{eq:lem1-1}
        P_k R( e_k 1_R - R H R)\psi = - P_k V R \psi = - \sum_{\abs{k-l}=1}P_k V P_l P_l R \psi
    \end{align}
    where we used $[R,H] = 0$ and $V = h M_x$. By assumption,
    \begin{align}\label{eq:lem1-2}
        e_k 1_R - R H R \geq (e_k - E) 1_R \geq (\Delta N - c_0)1_R
    \end{align}
    Combining \eqref{eq:lem1-1}, \eqref{eq:lem1-2}, and the triangle inequality, we have
    \begin{align}
        \norm{P_k R} \leq \frac{1}{\Delta N - c_0} \sum_{\abs{k-l}=1}\norm{P_k V P_l} \norm{P_l R}
    \end{align}
    Using $\norm{P_k V P_l} \leq h N$ and denoting $r_k \equiv \norm{P_k R}$, we find
    \begin{align}
        r_k \leq a(r_{k-1} + r_{k+1}),\quad a\equiv \frac{h N}{\Delta N - c_0}\in(0,1/2)
    \end{align}
    In the case of equality, the solution to $s_k = a (s_{k-1}+s_{k+1})$ is given by
    \begin{align}
        s_k = \lambda^{k - k_-+1} + \lambda^{k_+ - k+1},\quad \lambda  = \frac{1-\sqrt{1-4 a^2}}{2 a}
    \end{align}
    At the exterior boundary, we have $r_{k_--1}\leq 1\leq s_{k_--1}$ and $r_{k_++1}\leq 1\leq s_{k_++1}$. In the interior of the interval,
    we show $r_k\leq s_k$ by contradiction: suppose that, at $k$, $r_k - s_k = M$ attains its maximum. Then $s_k+M = r_k\leq a(r_{k-1}+r_{k+1})\leq a(s_{k-1}+s_{k+1}+2M) = s_{k} + 2 a M< s_k + M$.
\end{proof}

Applying Lemma \ref{lem: 1} to the low-energy subspace at $E = E_0 + \eta N$, with $E_0$ the ground-state energy of $H$, gives the following proposition:
\begin{prop}[Proposition 4 in the main text]\label{prop:1}
    Assume $2b(1-b)>2h+\eta$. For any $k\in I_b$, 
    \begin{align}
        \norm{P_k \Pi_{(-\infty, E_0+\eta N]}(H)} \leq C e^{-c d_{I_b}(k)}.
    \end{align}
\end{prop}
\begin{proof}
By the variational principle and the fact that the all-up state has zero expectation value of $H$, we have $E_0\leq 0$. 
    Using $e_k \geq 2b(1-b)N - c_0$ for $k\in I_b$, we have $e_k  - E \geq e_k - \eta N \geq (2b(1-b) - \eta)N - c_0 $. Applying Lemma \ref{lem: 1} with $\Delta \equiv 2b(1-b) - \eta$ (where $\Delta>2h$ by assumption), we have
    \begin{align}
        \norm{P_k \Pi_{(-\infty, E_0+\eta N]}(H)}\lesssim \lambda ^{k-bN+1} +  \lambda ^{(1-b)N-k+1} \leq 2 \lambda^{d_{I_b}(k)},
    \end{align}
    where $\lesssim$ means less than or equal to, up to a constant prefactor. Absorbing the constant prefactor into $C$ and denoting $c \equiv -\log\lambda>0$ completes the proof.
\end{proof}

Using Proposition \ref{prop:1}, we can estimate the thermal weight of the dynamical phases.
\begin{thm}\label{thm:static}
    Denoting $p_i = \tr(P_i \rho_\beta)$ for $i = \{A,B,C\}$, we have
    \begin{align}\label{eq:pa pb pc}
        p_A = p_C = \frac{1}{2} + O(\mathrm{poly}(N)e^{-c_B N}),\quad p_B\leq C_B \mathrm{poly}(N) e^{-c_B N}
    \end{align}
    Meanwhile, the initial leakage is exponentially small:
    \begin{align}\label{eq:leakage}
        \tr\left(P_C \frac{\rho_\beta^{1/2} P_A \rho_\beta^{1/2}}{p_A}\right)\leq C_{\mathrm{leak}}\mathrm{poly}(N) e^{-c_{\mathrm{leak}} N}
    \end{align}
\end{thm}
\begin{proof}
    Let $Q\equiv 1 - \Pi_{(-\infty, E_0+\eta N]}(H)$. Then
    \begin{align}
        \tr(Q \rho_\beta) = Z_\beta^{-1} \tr(Q e^{-\beta H})\leq e^{-N(\beta \eta  -\log 2)},
    \end{align}
    where we used $Z_\beta \geq 2^N e^{-\beta E_0}$. Let us choose $\eta$ so that $\beta\eta>\log 2$ and $2b(1-b)>2h+\eta$, which is always possible for large enough $\beta$ and small enough $h$. Hence, by energy splitting, 
    \begin{align}
        p_B = \tr(P_B \rho_\beta) & =\tr(P_B \Pi_{(-\infty, E_0+\eta N]}(H)\rho_\beta) + \tr(P_B Q \rho_\beta) \nonumber\\
        &\leq \norm{P_B \Pi_{(-\infty, E_0+\eta N]}(H)} + \tr(Q \rho_\beta)\nonumber\\
        & \leq \abs{B} C e^{-c(a-b)N } + e^{-N(\beta \eta  -\log 2)}
    \end{align}
    where we used H\"older's inequality in the first inequality. Choosing $c_B = \min\{c(a-b), \beta \eta  -\log 2\}$ and absorbing the constant prefactor into $C_B$, we arrive at \eqref{eq:pa pb pc}. Using the symmetry operator $U = \prod_i X_i$ and $U\rho_\beta U =\rho_\beta,\; U P_A U = P_C$, we have $p_A =p_C = (1-p_B)/2$.

    Next, we estimate the leakage by
    \begin{align}\label{eq: leak 1}
        \tr\left(P_C \frac{\rho_\beta^{1/2} P_A \rho_\beta^{1/2}}{p_A}\right) = p_A^{-1} \norm{P_C \rho_\beta^{1/2} P_A}_2^2\leq p_A^{-1} \norm{[ \rho_\beta^{1/2} ,P_A]}_2^2\leq \frac{\beta^2}{2 p_A} \norm{[H,P_A]\rho_\beta^{1/2}}_2^2
    \end{align}
    The first inequality follows from $P_C \rho_\beta^{1/2} P_A = P_C [\rho_\beta^{1/2},P_A] P_A$ and H\"older's inequality. The second inequality is obtained by a direct calculation in the eigenbasis:
    \begin{align}
        \norm{[ \rho_\beta^{1/2} ,P]}_2^2 = \sum_{mn}(\sqrt{p_m} - \sqrt{p_n})^2 \abs{P_{nm}}^2\leq \frac{1}{4} \sum_{mn} (\log p_m - \log p_n)^2(p_m+p_n) \abs{P_{nm}}^2 = \frac{\beta^2}{2} \norm{[H,P_A]\rho_\beta^{1/2}}_2^2
    \end{align} 
    Using the same energy-splitting argument as above for $p_B$, 
    \begin{align}
        \norm{[H,P_A]\rho_\beta^{1/2}}_2\leq \norm{[H,P_A]\Pi_{(-\infty, E_0+\eta N]}(H) } + \norm{[H,P_A]} \sqrt{\tr(Q\rho_\beta)}
    \end{align}
    Now, observe that the operator $[H,P_A]$ contains at most polynomially many bounded local terms and is nonzero only within an $O(1)$-width neighborhood of the cut $\partial A$. Hence, 
    \begin{align}
        \norm{[H,P_A]\rho_\beta^{1/2}}_2\lesssim C e^{-c(a-b)N } + \mathrm{poly}(N)e^{-N(\beta \eta  -\log 2)/2}
    \end{align}
 Using \eqref{eq: leak 1}, we arrive at \eqref{eq:leakage} by choosing $c_{\mathrm{leak}} = \min (2c(a-b),\beta \eta  -\log 2)$ and absorbing the constant prefactor into $C_{\mathrm{leak}}$.
\end{proof}

\subsection{A local operator injects finite energy}

Let $H = \sum_X h_X$.
Although the interaction graph is complete, the local energy strength is finite:
\begin{align}
    \Lambda:= \max_i \sum_{X \ni  i} \norm{h_X} = \frac{N-1}{N} + h <\infty
\end{align}
Therefore, a local operator cannot inject extensive energy. 
\begin{lem}\label{lem: energy inject}
 If a bounded local operator $\mathcal{O}$ is supported on most $\ell$ sites and $0<\tau<(2\Lambda)^{-1}$, then
 \begin{align}
     \norm{e^{\tau H} \mathcal{O} e^{-\tau H}} \leq \norm{\mathcal{O}} (1-2\Lambda \tau)^{-\ell}
 \end{align}
 Moreover, 
 \begin{align}
     \norm{\Pi_{[E+\delta E, \infty)}(H) \mathcal{O}\Pi_{(-\infty, E]}(H)} \leq \norm{\mathcal{O}} (1-2\Lambda \tau)^{-\ell} e^{-\tau \delta E}
 \end{align}
\end{lem}
\begin{proof}
    Expand
    \begin{align}
        \mathrm{ad}_H^m(\mathcal{O}) = \sum_{X_m,\ldots X_1}[h_{X_m},[\ldots [h_{X_1},\mathcal{O}]\ldots]] 
    \end{align}
    Since each interaction term has support on at most two sites, the accumulated support has a size of at most $\ell +m  -1$, yielding
    \begin{align}
        \norm{\mathrm{ad}_H^m(\mathcal{O})} \leq \norm{\mathcal{O}} (2\Lambda)^m \ell (\ell+1)\ldots (\ell+m-1)
    \end{align}
    Hence, for $2\Lambda\tau<1$, 
    \begin{align}
        \norm{e^{\tau H} \mathcal{O} e^{-\tau H}} \leq \sum_{m=0}^\infty \frac{\tau^m}{m!}\norm{\mathrm{ad}_H^m(\mathcal{O})} \leq  \norm{\mathcal{O}} \sum_{m=0}^\infty \frac{(2\Lambda\tau)^m}{m!} \ell (\ell+1)\ldots (\ell+m-1) =  \norm{\mathcal{O}} (1-2\Lambda \tau)^{-\ell}
    \end{align}
    Next, 
    \begin{align}
         \norm{\Pi_{[E+\delta E, \infty)}(H) \mathcal{O}\Pi_{(-\infty, E]}(H)}  &=  \norm{\Pi_{[E+\delta E, \infty)}(H) e^{-\tau H} e^{\tau H}\mathcal{O} e^{-\tau H} e^{\tau H}\Pi_{(-\infty, E]}(H)} \nonumber\\
         &\leq e^{-\tau (E+\delta E)} \cdot \norm{e^{\tau H} \mathcal{O} e^{\tau H}}\cdot e^{-\tau E}\nonumber\\
         &\leq \norm{\mathcal{O}} (1-2\Lambda \tau)^{-\ell} e^{-\tau \delta E}
    \end{align}
\end{proof}

Now, consider a $q$-seed word under separate real-time evolutions, $\mathcal{O} = S_q(t_q)\dots S_1(t_1)$, where each seed has a bounded norm $S_0$ and support size $s_0$. Choosing $\tau = 1/(4\Lambda)$, Lemma \ref{lem: energy inject} implies
\begin{align}
    \norm{e^{\tau H} \mathcal{O} e^{-\tau H}} \leq M_0^q,\quad M_0:= S_0 2^{s_0}
\end{align}
To see this, we insert $e^{\tau H} e^{-\tau H}$ between successive seeds and bound the imaginary-time evolution of each seed; the real-time evolution does not increase the norms. Moreover, we have
\begin{align}\label{eq:inject 1}
    \norm{\Pi_{[E+\delta E, \infty)}(H) \mathcal{O}\Pi_{(-\infty, E]}(H)} \leq M_0^q e^{-\delta E N/4\Lambda}.
\end{align}
This observation leads to the following lemma:
\begin{lem}[Lemma 6 in the main text]
    Consider the $q$-seed word $\mathcal{O}$ above. For a fixed $p>0$ and a fixed sufficiently large $\beta$, when $\abs{t_j}\leq \beta N^p$,
    \begin{align}
        \norm{[P_A,\mathcal{O}]\rho_\beta^{1/2}}_2 \leq C_{\mathrm{word}} \mathrm{poly}(N) e^{-c_{\mathrm{word}}N}.
    \end{align}
    The same holds for $P_C$.
\end{lem}
\begin{proof}
    Since $S_j(t_j) = U_j S_j U_j^\dagger, U_j = e^{\i H t_j}$, we can write $\mathcal{O}\equiv F_{3q} \dots F_1$ with each $F_j$ a seed or a unitary operator. Define $R_- := \Pi_{(-\infty, E+\eta_- N]}(H)$, $R_+ := \Pi_{(-\infty, E+\eta_+ N]}(H)$ with $0<\eta_-<\eta_+<2b(1-b)$ and $\beta\eta_->\log 2$ and $2b(1-b)>2h+\eta_+$. Set $Q_-:= 1-R_-,\; Q_+:= 1-R_+$. Using the energy-splitting argument, we have
    \begin{align}\label{eq:lem 9-1}
        \norm{[P_A,\mathcal{O}]\rho_\beta^{1/2}}_2 \leq \norm{[P_A,\mathcal{O}]R_-\rho_\beta^{1/2}}_2 + \norm{[P_A,\mathcal{O}]Q_-\rho_\beta^{1/2}}_2  
    \end{align}
    The second term can be bounded by the small thermal weight:
    \begin{align}\label{eq:lem 9-2}
        \norm{[P_A,\mathcal{O}]Q_-\rho_\beta^{1/2}}_2 \leq \norm{[P_A,\mathcal{O}]} \sqrt{\tr(Q_-\rho_\beta)} \leq 2 S_0^q e^{-N(\beta \eta_-  -\log 2)/2}.
    \end{align}
    To bound the first term, we first write the commutator as a telescoping sum:
    \begin{align}
        [P_A,\mathcal{O}] = \sum_{j=1}^{3q} L_j [P_A,F_j] W_j
    \end{align}
    with $L_j:= F_{3q}\dots F_{j+1}, W_j := F_{j-1}\dots F_1$. Then, we insert the higher-energy projector $R_+$ as follows:
    \begin{align}
        \norm{[P_A,\mathcal{O}]R_-} &\leq  \sum_{j=1}^{3q} \norm{L_j [P_A,F_j] R_+ W_j R_-} + \sum_{j=1}^{3q} \norm{L_j [P_A,F_j] Q_+ W_j R_-} \nonumber\\
        &\leq   \sum_{j=1}^{3q} \norm{L_j}\norm{[P_A,F_j]R_+} \norm{ W_j} \norm{P_-} + \sum_{j=1}^{3q} \norm{L_j} \norm{[P_A,F_j]} \norm{Q_+ W_j R_-} \nonumber\\
        &\leq C_{q,S_0} \sum_{j=1}^{3q} \left( \norm{Q_+ W_j R_-} + \norm{[P_A,F_j]R_+} \right),
    \end{align}
    where we used the fact that $\norm{L_j}, \norm{F_j},\norm{W_j}$ are bounded by constants that depend only on $q$ and $S_0$. The first term can be bounded using \eqref{eq:inject 1}. For the second term, there are two cases. If $F_j$ is a seed operator, $[P_A,S_j]$ has support near the cut $\partial A$ within a width of at most $s_0$ because $S_j$ can flip at most $s_0$ spins. According to Proposition \ref{prop:1}, $\norm{[P_A,S_j]R_+}\leq C_{s_0} e^{-c(a-b)N}  $. If $F_j$ is a unitary operator, 
    \begin{align}
        \norm{[P_A,U_j] R_+ } \leq  \int_0^{t_j} \norm{e^{\i H s} [P_A,H] e^{\i H(t_j-s)} R_+} \ud s \leq \abs{t_j} \norm{[P_A,H]R_+}\leq C \beta N^p e^{-c(a-b)N} ,
    \end{align}
    where we used $[H,R_+] = 0$. Combining these estimates, we arrive at
    \begin{align}\label{eq:lem 9-3}
        \norm{[P_A,\mathcal{O}]R_-\rho_\beta^{1/2}}_2\leq \norm{[P_A,\mathcal{O}]R_-} \leq  C_{q,S_0}  (3q)\left( M_0^q e^{-(\eta_+-\eta_-)N/4\Lambda} +  C_{s_0} e^{-c(a-b)N} +  C \beta N^p e^{-c(a-b)N} \right) \lesssim \mathrm{poly}(N)e^{-c'_{\mathrm{word}}N},
    \end{align}
    where $c'_{\mathrm{word}} = \min (c(a-b),(\eta_+-\eta_-))$. The proof is completed by combining \eqref{eq:lem 9-1}, \eqref{eq:lem 9-2}, and \eqref{eq:lem 9-3},
    choosing $c_{\mathrm{word}} = \min (c'_{\mathrm{word}},\beta \eta_-  -\log 2)$ and absorbing the constant prefactor into $C_{\mathrm{word}}$.
    The same clearly holds for $P_C$.
\end{proof}

\subsection{The Gaussian CKG sampler induces a small current}

The full expression for the CKG Lindbladian is
\begin{align}
    \mathcal{L}_{\mathrm{CKG}}(\rho) = -\i [B,\rho] + \sum_{a=1}^n\int_{\mathbb{R}} \mathcal{D}_{J_a(\omega)}(\rho)\ud \omega,
\end{align}
where $\mathcal{D}_L(\sigma)\equiv L \sigma L^\dagger - \frac{1}{2}\{L^\dagger L,\sigma\}$. 
The jump operator is given by
\begin{align}
    J_a(\omega) := \int_{\mathbb{R}} c_\beta(\omega,t)A_a(t)\ud t
\end{align}
with $c_\beta(\omega,t):= \sqrt{\gamma_\beta(\omega)/2\pi} f_\beta(t) e^{-\i\omega t}$ and
\begin{align}
    f_\beta(t)= \frac{e^{-t^2/\beta^2}}{\sqrt{\beta \sqrt{\pi/2}}},\quad \gamma_\beta(\omega) = \exp\left[ - \frac{(\beta\omega+1)^2}{2}\right].
\end{align}
The coherent term is given by 
\begin{align}
    B = \sum_{a=1}^{n} \int_{\mathbb{R}}b_1(u) e^{-\i\beta H u}\left( \int_{\mathbb{R}} b_2(v) A_a^\dagger(\beta v) A_a(-\beta v)\ud v \right)e^{\i\beta H u}\ud u
\end{align}

The Gaussian filter function allows us to approximate the jump operator by its finite-time component. In particular, let us truncate the time integral to $\abs{t}\leq \beta \Gamma$ and define 
\begin{align}
    \mathcal{L}_{\mathrm{CKG},\Gamma}(\rho) := -\i [B_\Gamma,\rho] + \sum_{a=1}^n\int_{\mathbb{R}} \mathcal{D}_{J_{a,\Gamma}(\omega)}(\rho)\ud \omega,
\end{align}
where 
\begin{align}
    J_{a,\Gamma}(\omega) &:= \int_{\abs{t}\leq \beta \Gamma} c_\beta(\omega,t)A_a(t)\ud t \nonumber \\
    B_\Gamma &:=\sum_{a=1}^n \int_{\abs{u}\leq \Gamma}b_1(u) e^{-\i\beta H u}\left( \int_{\abs{v}\leq \Gamma} b_2(v) A_a^\dagger(\beta v) A_a(-\beta v)\ud v \right)e^{\i\beta H u}\ud u
\end{align}
We have the following approximation:
\begin{prop}
    Suppose $n<\infty$, $\abs{\mathrm{supp} A_a}\leq a_0$ and $\norm{A_a}\leq A_0$.
    \begin{align}\label{eq: diamond 0}
        \norm{\mathcal{L}_{\mathrm{CKG}}-\mathcal{L}_{\mathrm{CKG},\Gamma}}_{\diamond} \leq 5 n_{\mathrm{jump}} A_0^2(e^{-\Gamma^2} + e^{-\Gamma})
    \end{align}
\end{prop}
\begin{proof}
Denote $T_{a,\Gamma}:= J_{a} - J_{a,\Gamma}$.
    Using the elementary diamond-norm estimate, we obtain 
    \begin{align}\label{eq: diamond 1}
        \norm{\mathcal{L}_{\mathrm{CKG}}-\mathcal{L}_{\mathrm{CKG},\Gamma}}_{\diamond} \leq \underbrace{\sum_a \int_{\mathbb{R}} \left(4\norm{J_{a,\Gamma}(\omega)}\norm{T_{a,\Gamma}(\omega)} +2 \norm{T_{a,\Gamma}(\omega)}^2 \right)\ud \omega }_{D}+ 2\norm{B - B_\Gamma}
    \end{align}
    Let us denote $m_\beta(\omega):= \int_{\mathbb{R}} \abs{c_\beta(\omega,t)}\ud t$. Observe that
    \begin{align}
        \frac{\int_{\abs{t}> \beta\Gamma} f_\beta(t)\ud t}{\int_{\mathbb{R}} f_\beta(t)\ud t}  = \mathrm{erfc}(\Gamma)
    \end{align}
    which leads to
    \begin{align}
        \norm{J_{a,\Gamma}(\omega)}&\leq A_0(1-\mathrm{erfc}(\Gamma))m_\beta(\omega) \nonumber\\
        \norm{T_{a,\Gamma}(\omega)}&\leq A_0\mathrm{erfc}(\Gamma)m_\beta(\omega)
    \end{align}
    Hence, 
    \begin{align}\label{eq: diamond 2}
        D \leq 4 n A_0^2 \mathrm{erfc}(\Gamma),
    \end{align}
    where we used $\int_{\mathbb{R}} m_\beta(\omega)^2\ud\omega = 1$.
    Next, denote $\mu_i:= \int_{\mathbb{R}} \abs{b_i(x)} \ud x$ and $\tau_i:= \int_{\abs{x}>\Gamma} \abs{b_i(x)} \ud x$. The explicit expressions for $b_i$ are \cite{ckg}
    \begin{align}
        b_1(u):=2 \sqrt{\pi} \mathrm{e}^{1 / 8}\left[\frac{1}{\cosh (2 \pi \cdot)} *\left(\sin (-\cdot) \mathrm{e}^{-2(\cdot)^2}\right)\right](u) ,\quad 
        b_2(u):= \frac{1}{\pi\sqrt{\pi}} e^{-4 u^2 - 2\i u}
    \end{align}
    Elementary analysis gives
    \begin{align}
        &\mu_1\leq \sqrt{\pi} \mathrm{e}^{1 / 8}/2,\quad \tau_1\leq 2 \sqrt{\pi} \mathrm{e}^{1 / 8}\left( \frac{1}{\pi} e^{-\pi \Gamma} + \frac{1}{4} e^{\Gamma^2/2}\right)\nonumber\\
        &\mu_2 = \frac{1}{2\pi},\quad \tau_2 = \frac{1}{2\pi}\mathrm{erfc}(2\Gamma)
    \end{align}
    Thus, 
    \begin{align}\label{eq: diamond 3}
        \norm{B - B_\Gamma}\leq n A_0^2(\mu_2 \tau_1+\mu_1 \tau_2)\leq n A_0^2\left(\frac{e^{1/8}}{\sqrt{\pi}} \left(\frac{1}{\pi} e^{-\pi \Gamma} + \frac{1}{4} e^{\Gamma^2/2}\right) + \frac{e^{1/8}}{4\sqrt{\pi}} \mathrm{erfc}(2\Gamma)\right)
    \end{align}
    Combining \eqref{eq: diamond 1}, \eqref{eq: diamond 2}, and \eqref{eq: diamond 3} and using $\mathrm{erfc}(x)\leq e^{-x^2}$, we arrive at \eqref{eq: diamond 0} by straightforward manipulation.
\end{proof}

Consequently, if we choose $\Gamma = N$, the current associated with the projection $P$ under the full dynamics can be approximated by the cutoff dynamics with an exponentially small error:
\begin{align}\label{eq:cutoff}
    \norm{\rho^{1/2} (\mathcal{L}^\dagger_{\mathrm{CKG}} - \mathcal{L}^\dagger_{\mathrm{CKG},N})(P) \rho^{1/2}}_1 \leq \norm{\mathcal{L}_{\mathrm{CKG}}-\mathcal{L}_{\mathrm{CKG},N}}_{\diamond}\lesssim \mathrm{poly}(N) e^{-N} ,
\end{align}
provided that the number of jump operators satisfies $n\lesssim \mathrm{poly}(N)$.

\begin{thm}
    Suppose $n\lesssim\mathrm{poly}(N)$, $\abs{\mathrm{supp} A_a}\leq a_0$ and $\norm{A_a}\leq A_0$. Then, for sufficiently large $\beta$ and small $h$, 
    \begin{align}
        \norm{\rho^{1/2}_\beta \mathcal{L}^\dagger_{\mathrm{CKG}}(P_C) \rho^{1/2}_\beta}_1 \leq C_J \mathrm{poly}(N) e^{-c_J N}
    \end{align}
\end{thm}
\begin{proof}
    Using \eqref{eq:cutoff}, we only need to estimate the cutoff dynamics. Denote $ \mathcal{L}_{D,N}:= \sum_{a}\int_{\mathbb{R}} \mathcal{D}_{J_{a,N}(\omega)}(\rho)\ud \omega$. Then
    \begin{align}
        \norm{\rho^{1/2}_\beta \mathcal{L}_{D,N}^\dagger(P_C) \rho^{1/2}_\beta}_1 &\leq \sum_{a}\int_{\mathbb{R}} \norm{J_{a,N} \rho_\beta^{1/2}}_2 \norm{[P_C,J_{a,N}] \rho_\beta^{1/2}}_2 \ud\omega \nonumber\\
        & \leq \left(\left(\sum_{a}\int_{\mathbb{R}} \norm{J_{a,N}}^2 \ud\omega\right) \left(\sum_{a}\int_{\mathbb{R}} \norm{[P_C,J_{a,N}]\rho_\beta^{1/2}}_2^2 \ud\omega\right)\right)^{1/2}\nonumber\\
        &\leq \left(\left(n A_0^2\right)  \left(n   A_0^2 \sup_{a,\abs{t}\leq \beta N} \norm{[P_C,A_{a}(t)]\rho_\beta^{1/2}}_2^2 \right)\right)^{1/2}
    \end{align}
    where we used $\int_{\mathbb{R}} (\int_{\mathbb{R}} \abs{c_\beta(\omega,t)}\ud t)^2\ud\omega = 1$. Applying Lemma \ref{lem: seed} with $\mathcal{O} = A_a(t)$, we have 
    \begin{align}\label{eq:current 1}
        \norm{\rho^{1/2}_\beta \mathcal{L}_{D,N}^\dagger(P_C) \rho^{1/2}_\beta}_1 \leq n A_0^2 C_{\mathrm{word}} \mathrm{poly}(N) e^{-c_{\mathrm{word}}N}
    \end{align}
    Next, the coherent part satisfies
    \begin{align}\label{eq:current 2}
        \norm{\rho^{1/2}_\beta [B_N,P_C] \rho^{1/2}_\beta}_1 &\leq \norm{[B_N,P_C] \rho^{1/2}_\beta}_2 \nonumber\\
        &\leq n\int_{\mathbb{R}}\abs{b_1(u)}\ud u \int_{\mathbb{R}}\abs{b_2(u)}\ud u \sup_{a,\abs{u},\abs{v}\leq \beta N} \norm{[P_C,M_a(u,v)]\rho_\beta^{1/2}}_2 \nonumber\\
        &\leq n \frac{e^{1/8}}{4\sqrt{\pi}}C_{\mathrm{word}} \mathrm{poly}(N) e^{-c_{\mathrm{word}}N},
    \end{align}
    where $M_a(u,v):= e^{-\i \beta H u }A_a^\dagger(\beta v) A_a(-\beta v) e^{\i \beta H u }$, and in the last inequality we applied Lemma \ref{lem: seed}. Therefore, combining \eqref{eq:cutoff}, \eqref{eq:current 1}, and \eqref{eq:current 2} completes the proof. 
\end{proof}

\section{An effective Fokker-Planck equation}\label{app: mean field}
In this section, we derive an effective Fokker-Planck equation from the Lindbladian dynamics in the large-$N$ limit.

Following section 4 in the main text, let us write $r:= \sqrt{h^2+x^2}$ and introduce the change of variables
\begin{align}
    S:= \frac{h X + m Z}{r},\quad T:= \frac{m X - h Z}{r}
\end{align}
with $\{S,T\} = 0,\; [S,T]=\i Y$. Choose the seed jump operators to be $A_S:= S/\sqrt{3},A_T:= T/\sqrt{3},A_Y:= Y/\sqrt{3} $. Define the ladder operator $L:= (T+\i Y)/2$ with $[H_{\mathrm{MF}},L] = -2 r L$. A straightforward calculation gives
\begin{align}
    J_S(\omega) &= \frac{\gamma_\beta(\omega)}{3} \phi_\beta(\omega)S,\nonumber\\
    J_T(\omega) & = \frac{\gamma_\beta(\omega)}{3}\left(\phi_\beta(\omega+2r) L + \phi_\beta(\omega-2 r) L^\dagger\right), \nonumber\\
    J_Y(\omega)& = \frac{\gamma_\beta(\omega)}{3} \left(-\i \phi_\beta(\omega+2r)L+\i \phi_\beta(\omega-2r)L^\dagger\right),
\end{align}
where
\begin{align}
    \phi_\beta(x):= \frac{1}{\sqrt{2\pi}}\int \ud s f_\beta(s) e^{-\i x s} = \sqrt{ \frac{\beta}{\sqrt{2\pi}}} e^{-\beta^2 x^2/4}
\end{align}
Substituting this into the expression for the dissipative generator, we obtain
\begin{align}
    \LL_{D} = \gamma_\downarrow \mathcal{D}_L + \gamma_\uparrow \mathcal{D}_{L^\dagger}+\gamma_0 \mathcal{D}_S
\end{align}
where the rates are given by
\begin{align}
    \gamma_\downarrow &= \frac{2}{3\sqrt{2}} e^{-(1-2\beta r)/4},\quad \gamma_\uparrow = \frac{2}{3\sqrt{2}} e^{-(1+2\beta r)/4} ,\quad \gamma_0 = \frac{e^{-1/4}}{3\sqrt{2}} 
\end{align}
Note that $G:= \sum_a \int \ud\omega~ J_a^\dagger(\omega) J_a(\omega) = (\gamma_0 + (\gamma_\uparrow+\gamma_\downarrow)/2)I - ((\gamma_\uparrow-\gamma_\downarrow)/2) S $ commutes with the mean-field Hamiltonian, $[G,H_{\mathrm{MF}}] = 0$, so the coherent term is found to vanish:
\begin{align}
    B_{\mathrm{CKG,MF}}= \frac{\i}{2}\tanh\left(\frac{\beta}{4} \mathrm{ad}_{H_{\mathrm{MF}}}\right) G = 0
\end{align}
where the expression is proved in \cite{ckg}.
Applying the full Lindbladian generator to a general two-level state $\rho(t) = \frac{1}{2}(I+x(t)X+y(t)Y+z(t)Z)$, we have
\begin{align}
    \dot u &= (\gamma_\uparrow-\gamma_\downarrow) - (\gamma_\uparrow+\gamma_\downarrow)u \nonumber\\
    \dot v &= - \gamma_\perp v,\nonumber\\
    \dot y &= -\gamma_\perp y,
\end{align}
with $\gamma_\perp:= (\Gamma_\uparrow+\gamma_\downarrow)/2+2\gamma_0$ and $u:= (h x+m z)/r,\; v:=(m x - h z)/r$.
Consider dissipation near the saddle point $(x,y,z) = (x_s,0,0)$ with $x_s = \tanh(\beta h)$. We find
\begin{align}\label{eq: eom saddle}
    \delta\dot{ x} = -\gamma_x \delta x,\quad \delta\dot{ y} = -\gamma_y \delta y,\quad \delta\dot{ z} = \gamma_z \delta z
\end{align}
where
\begin{align}
    \gamma_x &: = \frac{4e^{-1/4}}{3\sqrt{2}} e^{-(\beta h)^2} \cosh(\beta h) \nonumber\\
    \gamma_y &: = \frac{4e^{-1/4}}{6\sqrt{2}} \left( 1+ e^{-(\beta h)^2} \cosh(\beta h) \right) \nonumber\\
    \gamma_z &: = \frac{4e^{-1/4}}{3\sqrt{2}h} \left( 1+ e^{-(\beta h)^2} \cosh(\beta h) \right) (\tanh(\beta h) - h)
\end{align}
% For sufficiently large $\beta$, $\tanh(\beta h) > h$, so $\gamma_m>0$ indicates that the $z$-direction is unstable. 

To capture the fluctuations near the saddle point, we need to retain the leading $1/N$ contributions. Set
\begin{align}
    F:= \frac{1}{N} \sum_i(\mathcal{O}_i - \langle \mathcal{O}_i \rangle)
\end{align}
In the large-$N$ limit, we have
\begin{align}
    \norm{[J_a,F]} = O(N^{-1}),\quad \norm{[[J_a,F],F]} = O(N^{-2}),\quad \norm{[\mathcal{L}^\dagger F,F]} = O(N^{-2})
\end{align}
where we used $[\mathcal{L}^\dagger F,F] \sim [F,F]=0$ from the analysis of the infinite-$N$ limit above; \eqref{eq: eom saddle} is diagonal. Hence, based on Appendix~\ref{app:FP}, any function of $F$ obeys 
\begin{align}\label{eq:FP 1}
    \LL^\dagger f(F)=  (\LL^\dagger F) f'(F) + \frac{1}{2}\Gamma(F) f''(F) +R(f),\quad \norm{R(f)}\leq O(N^{-2})
\end{align}
where
\begin{align}
    \Gamma(F):= \sum_a C_a^\dagger C_a = \sum_a [J_a,F]^\dagger[J_a,F] 
\end{align}
Note that for $O(N)$ channels, $\norm{\Gamma(F)} = O(N^{-1})$. Taking the observable to be the local magnetization $F =  m = \frac{1}{N} \sum_i Z_i$ and evaluating \eqref{eq:FP 1} at the infinite-$N$ saddle point, we obtain an effective (dual) Fokker-Planck equation
\begin{align}
    \mathcal{L}^\dagger f( m) = \gamma_s m f'( m) + \frac{D}{N} f''( m)
\end{align}
with $\gamma_s\approx \gamma_z$ in the $h\to 0$ limit and $D:= \frac{N}{2} \langle \Gamma(F) \rangle_{\mathrm{MF}}$. By detailed balance, the value of $\gamma_s$ is determined by $D$ and the equilibrium distribution at the saddle point
\begin{align}
    \gamma_s = \frac{D}{N} \frac{\ud^2}{\ud m^2} \log p_{\mathrm{eq}}(m)|_{m=0}  \approx \beta D,
\end{align}
where we assumed large $\beta$ and small $h$. This can be seen by noting that $p_{\mathrm{eq}}(m) \propto \binom{N}{N(1+m) / 2} e^{\beta N m^2/2} \approx e^{\beta N m^2/2}$ for $m\ll 1$ and $\beta\gg 1$.
We omit the explicit expressions for $\gamma_s,D$; what is important is that both are positive. The corresponding Fokker-Planck equation for the distribution $p=p(m,t)$ is given by
\begin{align}
    \partial_t p = - \partial_m(\gamma_s  m p)+ \frac{D}{N} \partial_m^2 p 
\end{align}
which takes the same form as that in \cite{gwl}.

\subsection{From Lindblad equations to diffusive Fokker-Planck equations}\label{app:FP}

Consider, for Hermitian $F$,
\begin{align}
    K:= \mathcal{L}^\dagger F = \i[B,F]+\sum_a(J_a^\dagger F J_a - \frac{1}{2}\{J_a^\dagger J_a,F\})
\end{align}
Let
\begin{align}
    \Gamma_{\mathcal{L}}(F):= \mathcal{L}^\dagger(F^2) - (\mathcal{L}^\dagger F)F - F(\mathcal{L}^\dagger F).
\end{align}
Substituting the expression for $\mathcal{L}^\dagger$, we obtain
\begin{align}
    \Gamma_{\mathcal{L}}(F):= \sum_a C_a^\dagger C_a = \sum_a [J_a,F]^\dagger[J_a,F] \geq 0
\end{align}
Take $U_u:= e^{\i u F}$. Then
\begin{align}
    -\i \partial_u(\mathcal{L}^\dagger U_u) = \mathcal{L}^\dagger(F U_u) = K U_u +  F \mathcal{L}^\dagger(U_u) + \sum_a C_a^\dagger \left(\i \int_0^u e^{\i(u-w)F} C_a e^{\i w F}\ud w\right),
\end{align}
where we used the Duhamel formula and $C_a:= [J_a,F]$. Integrating over $u$, we obtain
\begin{align}
    \mathcal{L}^\dagger U_u = \underbrace{\i\int_0^u e^{\i(u-v)F} K e^{\i v F}\ud v}_{I_1} - \underbrace{\sum_a  \int_0^u \ud v  \int_0^v \ud w ~ e^{\i(u-v)F} C_a^\dagger e^{\i(v-w)F} C_a e^{\i w F}}_{I_2}
\end{align}
Commuting the operators, we arrive at
\begin{align}
    I_1 &= \i u U_u K + R_1(u),\nonumber\\
    I_2 &= \frac{u^2}{2} U_u \Gamma_\LL(F) + R_2(u),
\end{align}
where
\begin{align}
    R_1(u)&:= \i \int_0^u e^{\i (u-v)F}[K,e^{\i v F}]\ud v ,\nonumber\\
    R_2(u)&:=\sum_a  \int_0^u \ud v  \int_0^v \ud w ~e^{\i(u-v)F} C_a^\dagger e^{\i(v-w)F} [C_a, e^{\i w F}]+ \sum_a  \int_0^u v \ud v ~  e^{\i(u-v)F} [C_a^\dagger, e^{\i v F}] C_a 
\end{align}
The operator norms of the residual terms are bounded by
\begin{align}
    \norm{R_1(u)} &\leq \int_0^u v \ud v \norm{[K,F]} =: \frac{u^2}{2} \epsilon_1,\nonumber\\
    \norm{R_2(u)} &\leq \sum_a  \int_0^u \ud v  \int_0^v w\ud w \norm{C_a} \norm{[C_a,F]}+ \sum_a  \int_0^u v^2\ud v\norm{C_a} \norm{[C_a,F]}=: \frac{u^3}{2} \epsilon_2
\end{align}
Combining these bounds gives
\begin{align}\label{eq:app FP 1}
    \norm{\LL^\dagger U_u - \i u U_u K + \frac{u^2}{2}U_u\Gamma_\LL(F)}\leq \frac{u^2}{2} \epsilon_1 + \frac{u^3}{2} \epsilon_2 
\end{align}
Now, define a function of the operator $F$ by $f(F) = \int_\mathbb{R} \hat{f}(u) e^{\i u F} \ud u$ and set $M_k(f):= \int \abs{u}^k \abs{\hat{f}(u)}
\ud u <\infty$. Then \eqref{eq:app FP 1} implies
\begin{align}
    \LL^\dagger f(F)= K f'(F) + \frac{1}{2}\Gamma_\LL(F) f''(F) +R(f)
\end{align}
with 
\begin{align}
    \norm{R(f)}\leq \frac{1}{2}M_2(f)\epsilon_1 + \frac{1}{2}M_3(f)\epsilon_2
\end{align}

\end{document}